\documentclass[aps,prl,twocolumn,groupedaddress]{revtex4}
\usepackage[dvips]{graphicx}

\begin{document}
\title{Scaling properties of granular rheology near the jamming transition}
\author{Takahiro Hatano}
\affiliation{Earthquake Research Institute, University of Tokyo, 
1-1-1 Yayoi, Bunkyo, Tokyo 113-0032, Japan}
\date{\today}

\begin{abstract}
Rheological properties of a dense granular material consisting of frictionless spheres are investigated.
It is found that the shear stress, the pressure, and the kinetic temperature obey critical scaling 
near the jamming transition point, which is considered as a critical point.
These scaling laws have some peculiar properties in view of conventional critical phenomena 
because the exponents depend on the interparticle force models so that they are not universal.
It is also found that these scaling laws imply the relation between 
the exponents that describe the growing correlation length.
\end{abstract}

\maketitle

Jamming is a general concept that a disordered system cannot flow under nonzero driving force 
(e.g., shear stress), which is generally realized at higher densities and lower temperatures.
Particularly, in a class of athermal (i.e., zero temperature) systems 
such as emulsions, foams, and frictionless granular particles, 
the shear modulus and the bulk modulus emerge/vanish above/below a definite density 
\cite{liu,ohern,ohern2,zhang}.
This rigidity transition is now referred to as the jamming transition and 
the transition point is referred to as {\it point J}.
Thanks to massive experiments and numerical simulations which indicate 
the growing correlation length and characteristic time \cite{drocco,dauchot,lechenault,durian1,durian2}, 
it seems to be confirmed that point J is a critical point, although the critical fluctuation is not observed 
in static quantities but found in purely dynamic quantities such as the particle displacement in a certain time window.
Quite interestingly, the critical fluctuation in dynamic quantities is also observed in 
supercooled liquids \cite{yamamoto,berthier} so that glass transitions might be related to the criticality of point J.
Therefore, clarification of the critical nature of point J is important 
in the context of glass physics as well as granular mechanics.

As a result of the criticality of point J, Olsson and Teitel find a scaling law for rheology \cite{olsson}, 
which is of the same form as those in conventional critical phenomena.
Such scaling laws provide us a systematic way to estimate critical exponents, 
which are of special importance in the context of the universality of critical phenomena.
With regard to the jamming transition, however, it is known that some exponents 
that describe the elastic moduli depend on the force model of particles \cite{ohern2} 
so that the universality in the sense of conventional critical phenomena is questioned.
Along the line of thought, the model dependence of the critical exponents must be further investigated via scaling laws.
In this paper, several scaling laws for rheological properties of a granular material are investigated 
by means of molecular dynamics simulation.
It is found that these scaling laws are quite different from those for conventional critical phenomena 
in the sense that the exponents depend on the interparticle force models.
We then briefly discuss the relation between these exponents and the growing correlation length.

We consider a bidisperse mixture of frictionless particles, 
the diameters of which are $d$ and $0.7d$, respectively.
The diameter and the position of particle $i$ are denoted by $R_i$ and ${\bf r}_{i}$, respectively.
The force between particles $i$ and $j$ is written as 
${\bf F}_{ij}= \left[kF(\delta_{ij})+\zeta{\bf n}_{ij}\cdot\dot{{\bf r}}_{ij}\right]{\bf n}_{ij}$, 
where ${\bf n}_{ij}={\bf r}_{ij}/|{\bf r}_{ij}|$, ${\bf r}_{ij}={\bf r}_{i}-{\bf r}_{j}$, 
and $\delta_{ij}$ denotes the overlap length, $(R_i+R_j)-|{\bf r}_{ij}|$.
(Note that $\delta_{ij}=0$ if $(R_i+R_j)<|{\bf r}_{ij}|$).
Here we test two types of interactions: $F(z)=z$ (the linear force model) 
and $F(z)=z^{1.5}$ (the Hertzian force model).
Throughout this study, we adopt the units in which $d=1$, $M=1$, and $\zeta=1$.
Here $M$ denotes the mass of the larger particles, while that of the smaller particles is $0.7^3M$.
We consider sufficiently hard spheres; $k=10^2$ for the linear force model 
and $k=10^4$ for the Hertzian force model.

In order to ensure uniform shear flow, we adopt the SLLOD equations 
together with the Lees-Edwards boundary conditions \cite{allen}.
\begin{eqnarray}
{\dot {\bf q}}_i &=& \frac{{\bf p}_i}{m_i} + \gamma q_{z,i} {\bf n}_y,\\
{\dot {\bf p}}_i &=& \sum_{j}{\bf F}_{\rm ij} - \gamma p_{z,i} {\bf n}_y, 
\end{eqnarray}
where $\gamma$ denotes the shear rate.
The system is of constant volume and consists of $10^3$ particles \cite{finitesize}.
Here we investigate several volume densities ranging from $0.5$ to $0.7$.
We discuss the rheological properties of this system with respect to two control parameters: 
the shear rate $\gamma$ and the volume density $\phi$.
Note that the shear stress and the pressure is defined through the virial \cite{allen}.

\begin{figure}[tb]
\begin{center}
\includegraphics[scale=0.35]{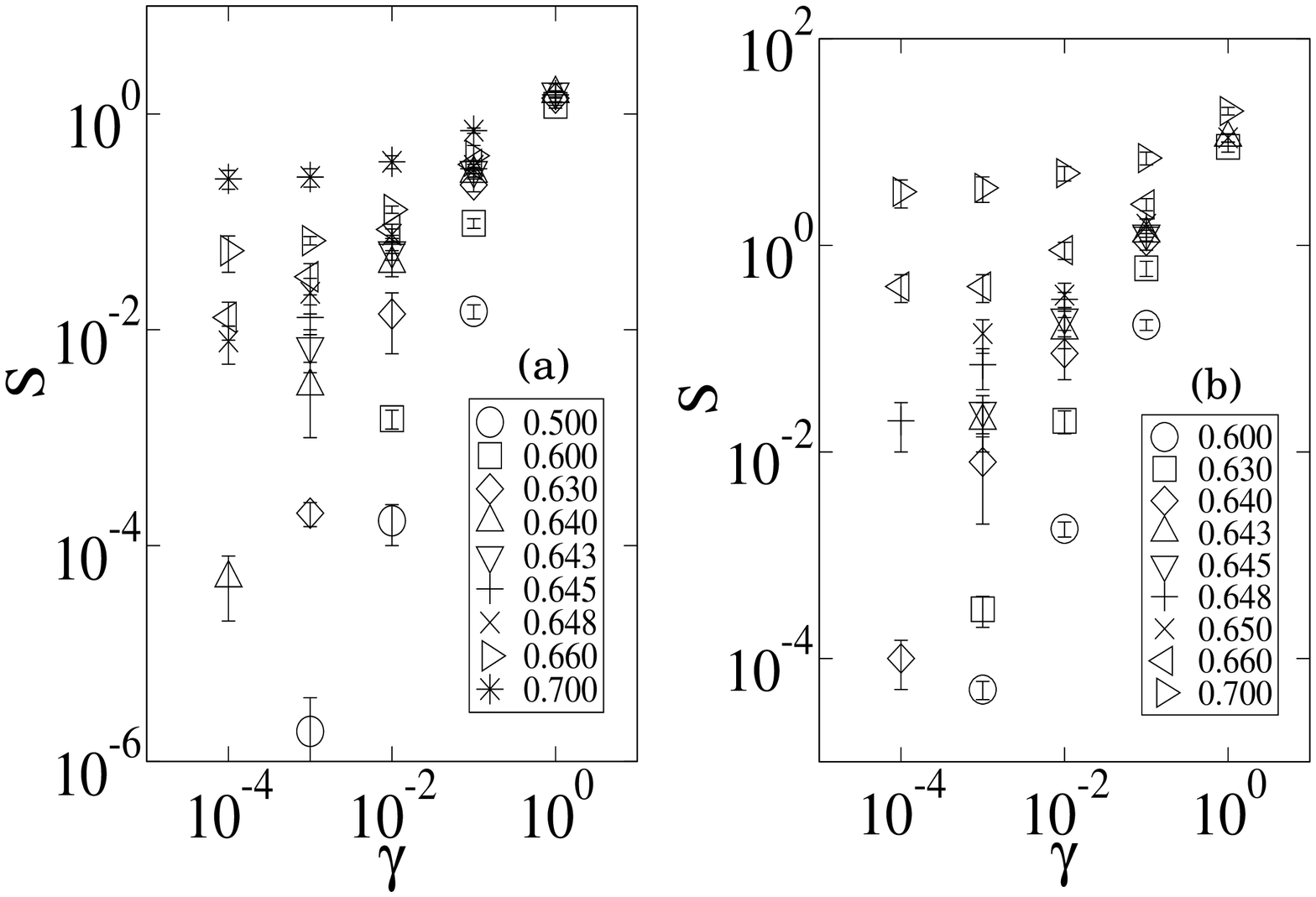}
\end{center}
\caption{\label{shearstress} The shear rate dependence of the shear stress of granular materials 
subjected to stationary shear flow at several densities.
(a) The model in which the particles interact via linear repulsive force. (The linear force model).
(b) The model in which the particles interact via Hertzian force. (The Hertzian force model).}
\end{figure}

First we discuss the shear rate dependence of the shear stress. 
As is shown in Fig. \ref{shearstress}, the linear model and the Hertzian model 
show qualitatively the same behaviors, albeit quantitatively different.
At lower densities, the shear stress is proportional to the square of the shear rate,  
which is known as the Bagnold's scaling \cite{silbert1,silbert2}.
At higher density, the shear stress seems to have a nonzero value in the zero shear rate limit, 
which is interpreted as the yield stress.
From these data, it is found that the threshold density $\phi_J$ (i.e., the jamming density) 
is somewhere between $0.645$ and $0.648$ in the present system.

\begin{figure}[tb]
\begin{center}
\includegraphics[scale=0.35]{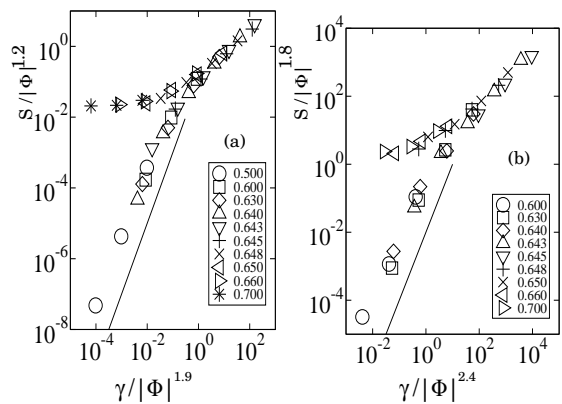}
\end{center}
\caption{\label{Sscaled} Collapse of the rheological data show in Fig. \ref{shearstress} 
using the scaling law, Eq. (\ref{scalingS}).
The solid lines are proportional to $\gamma^2$, which represent the Bagnold's scaling.
(a) The scaling law Eq. (\ref{scalingS}) for the linear force model.
(b) The scaling law Eq. (\ref{scalingS}) for the Hertzian force model.}
\end{figure}

As is shown in Fig. \ref{Sscaled}, these rheological data are collapsed using the following relation.
\begin{equation}
\label{scalingS}
S = |\Phi|^{y_{\Phi}}{\cal S}_{\pm}\left(\frac{\gamma}{|\Phi|^{y_{\Phi}/y_{\gamma}}}\right),
\end{equation}
where ${\cal S}_{\pm}(\cdot)$ is the scaling function and $\Phi$ denotes $\phi_J-\phi$.
The jamming density $\phi_J$ is set to be $0.646$, which we use throughout 
the present numerical analysis \cite{note_phiJ}.
Note that this scaling law, Eq. (\ref{scalingS}), is of the same form as those in conventional critical phenomena.
One may thus expect that the jamming transition is one of the conventional critical phenomena 
and that the exponents $y_{\Phi}$ and $y_{\gamma}$, together with the scaling dimension of 
the shear stress, classify the jamming transition into a universality class.
However, this idea is ruled out because these exponents depend on the force model.
We obtain $y_{\Phi}=1.2\pm0.1$ and $y_{\Phi}/y_{\gamma}=1.9\pm 0.1$ for the linear force model, 
while $y_{\Phi}=1.8\pm0.1$ and $y_{\Phi}/y_{\gamma}=2.4\pm 0.1$ for the Hertzian force model.
The exponent $y_{\gamma}$ is thus roughly estimated as $0.63$ and $0.75$ 
for the linear model and for the Hertzian model, respectively.
Recalling that the elastic moduli of granular media depends on the interaction force \cite{ohern2}, 
the model dependence of $y_{\Phi}$ is rather obvious in view of Eq. (\ref{yieldstress}).
Furthermore, note that the other exponent $y_{\gamma}$ also depends on the force model.
These exponents are listed in Table. \ref{exponents}.

\begin{table}
\caption{\label{exponents}
The exponents estimated from scaling laws, Eqs. (\ref{scalingS}), (\ref{scalingP}), and (\ref{scalingT}). }
\begin{center}
\begin{tabular}{l|cccccc} \hline
Observable & \multicolumn{2}{c}{Shear stress} &  \multicolumn{2}{c}{Pressure} & 
\multicolumn{2}{c}{Temperature}  \\ \hline
Exponent & $y_{\Phi}$ & $y_{\Phi}/y_{\gamma}$ & $y'_{\Phi}$ & $y'_{\Phi}/y'_{\gamma}$
& $x_{\Phi}$ & $x_{\Phi}/x_{\gamma}$ \\ \hline
Linear force& 1.2(1) & 1.9(1) & 1.2(1) & 2.1(2) & 2.5(2) & 1.9(2) \\
Hertzian force &  1.8(1) & 2.4(1) & 1.8(1) & 2.4 (1) & 3.2(2) & 2.4(2)\\ \hline
\end{tabular}
\end{center}
\end{table}

The scaling law Eq. (\ref{scalingS}) implies two nontrivial rheological properties.
First, it describes power-law rheology at $\Phi=0$, where the right hand side of 
Eq. (\ref{scalingS}) does not depend on $\Phi$.
This leads to $S\propto\gamma^{y_{\gamma}}$.
Note that the exponent $y_{\gamma}\simeq 0.63$ for the linear force model 
is directly observed in the previous studies \cite{hatano1,hatano2}.
Second, Eq. (\ref{scalingS}) also describes the density dependence of the dynamic yield stress.
Inserting $\gamma = 0$ into Eq. (\ref{scalingS}), we get 
\begin{equation}
\label{yieldstress}
S(\Phi, 0)={\cal S}_{-}(0)|\Phi|^{y_{\Phi}}.
\end{equation}
The dynamic yield stress is thus proportional to $|\Phi|^{y_{\Phi}}$, 
provided that ${\cal S}_{-}(0)\neq0$.

Then we discuss the behavior of the pressure (i.e., the normal stress).
One of the essential features of granular rheology is that the pressure 
also depends on the shear rate \cite{hatano1}, 
because granular systems have no intrinsic temperature 
and the extent of the random motion of particles increases with the shear rate.
Remarkably, they also obey a scaling law; 
\begin{equation}
\label{scalingP}
P = |\Phi|^{y'_{\Phi}}{\cal P}_{\pm}\left(\frac{\gamma}{|\Phi|^{y'_{\Phi}/y'_{\gamma}}}\right),
\end{equation}
where ${\cal P}_{\pm}(\cdot)$ is the scaling function.
The exponents are estimated as $y'_{\Phi}=1.2\pm0.1$ and $y'_{\Phi}/y'_{\gamma}=2.1\pm0.2$ 
for the linear force model, and $y'_{\Phi}=1.8\pm0.1$ and $y'_{\Phi}/y'_{\gamma}=2.4\pm0.1$ 
for the Hertzian force model.
The exponent $y'_{\gamma}$ for each model is thus roughly estimated as $0.57$ and $0.72$, respectively.
Note that the exponents again depend on the force model.
\begin{figure}[tb]
\begin{center}
\includegraphics[scale=0.35]{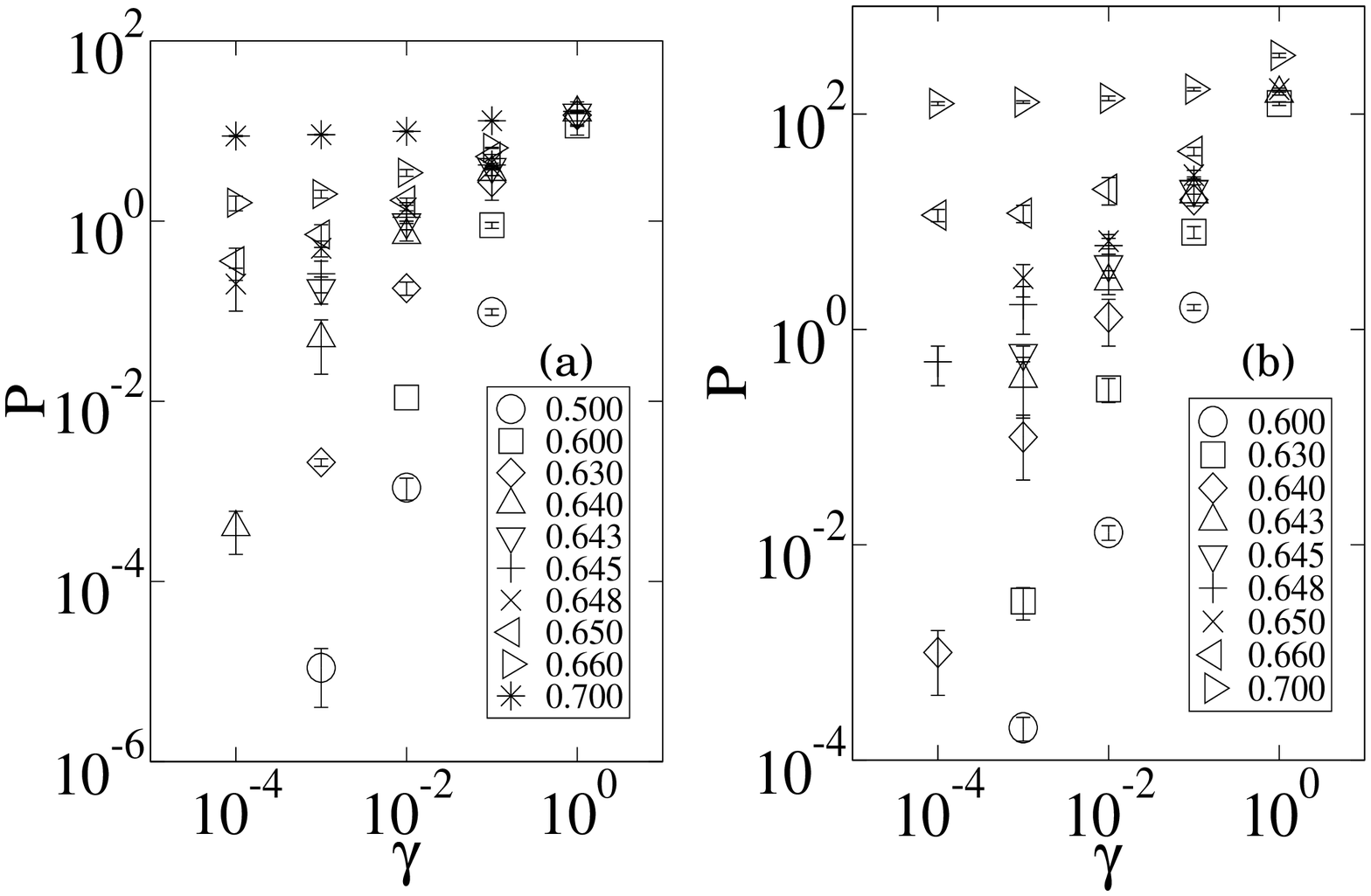}
\end{center}
\caption{\label{pressure} The shear rate dependence of the pressure 
(i.e., the normal stress $P_{zz}$) of granular materials subjected to stationary shear flow at several densities.
(a) The linear force model. (b) The Hertzian force model}.
\end{figure}

This scaling law, Eq. (\ref{scalingP}), again implies two important rheological behaviors.
First, inserting $\gamma=0$ into Eq. (\ref{scalingP}) leads to the density dependence 
of the pressure in the $\gamma\rightarrow0$ limit; 
$P(\Phi, 0)\propto |\Phi|^{y'_{\Phi}}$.
We remark that this pressure is {\it the dynamic yield pressure} 
and thus should not be identified with the pressure at zero shear strain, 
which is proportional to $|\Phi|^{1.0}$ for the linear force model 
and $|\Phi|^{1.5}$ for the Hertzian force model \cite{ohern,ohern2}.
Second, inserting $\Phi=0$ into Eq. (\ref{scalingP}) leads to the power-law behavior 
at the jamming density; $P(0, \gamma) \propto \gamma^{y'_{\gamma}}$.
Note that $y'_{\gamma}\simeq0.57$ for the linear force model is consistent with 
the previous study \cite{hatano1,hatano2}.
\begin{figure}[tb]
\begin{center}
\includegraphics[scale=0.35]{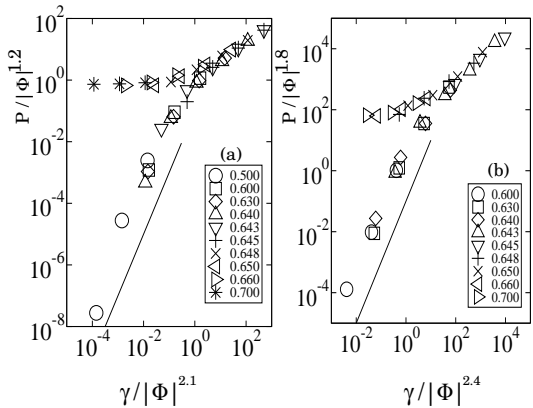}
\end{center}
\caption{\label{Pscaled} Collapse of the rheological data shown in Fig. \ref{pressure} 
using the scaling law, Eq. (\ref{scalingP}).
The dashed lines are proportional to $\gamma^2$, which represent the Bagnold's scaling.
(a) The scaling law for the linear force model.
(b) The scaling law for the Hertzian force model.}
\end{figure}

As is discussed above, the exponents in the scaling laws depend on a force model.
This is partially understood if one recalls that $y_{\Phi}$ and $y'_{\Phi}$ describe 
the dynamic yield stress and the dynamic yield pressure, respectively.
These quantities are essentially dominated by the elastic moduli of the system, 
the exponents for which do depend on the force model \cite{ohern2}.
Along the line of thought, dynamical quantities that vanish in the static limit may have 
a universal scaling behavior which does not depend on the force model.
To test this idea, we discuss the scaling properties of the kinetic temperature defined by 
$T=\langle m_i\delta {\bf v}_i^2\rangle/3$.
Note that this kinetic temperature is different from several effective temperatures 
that have been tested in sheared athermal systems, which take a nonzero value in the static limit \cite{ono}.
These effective temperatures are dominated by the static limit value and hence insensitive to the shear rate, 
while the kinetic temperature is much more sensitive to the dynamics.
The scaling law for the kinetic temperature reads 
\begin{equation}
\label{scalingT}
T(\Phi, \gamma)=|\Phi|^{x_{\Phi}}{\cal T}_{\pm}\left(\frac{\gamma}{|\Phi|^{x_{\Phi}/x_{\gamma}}}\right),
\end{equation}
where ${\cal T}_{\pm}(\cdot)$ are the scaling functions.
We stress that the exponents again depend on the force model: 
$x_{\Phi}=2.5\pm 0.2$ and $x_{\Phi}/x_{\gamma}=1.9\pm 0.2$ for the linear force model 
and $x_{\Phi}=3.2\pm 0.2$ and $x_{\Phi}/x_{\gamma}=2.4\pm 0.2$ for the Hertzian force model.
Therefore, the exponents for the kinetic temperature, which is a purely dynamic quantity, 
do depend on the force model.
Therefore we can conclude that the exponents in the rheological scaling laws are not universal 
in contrast to conventional critical phenomena.
\begin{figure}[tb]
\begin{center}
\includegraphics[scale=0.35]{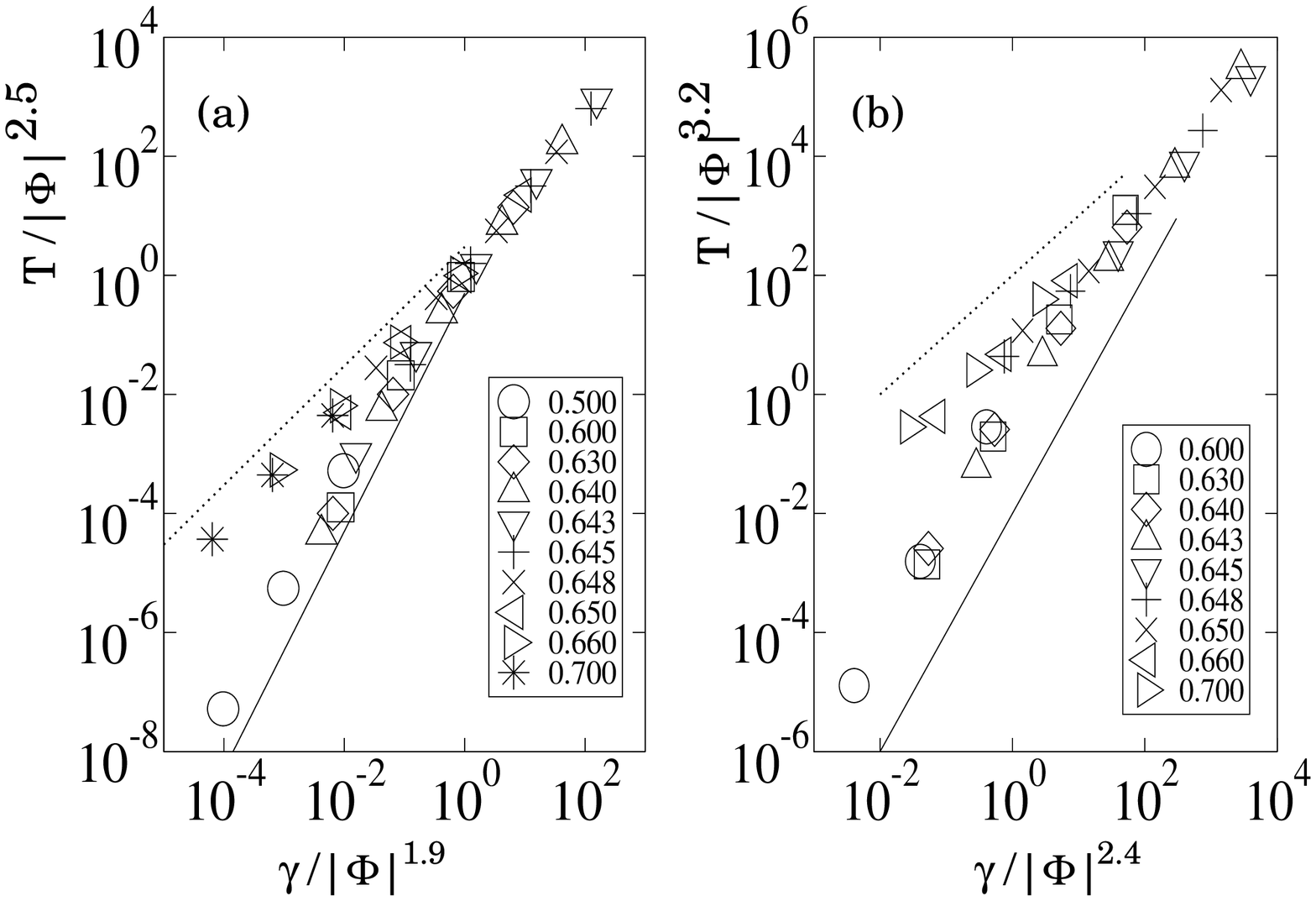}
\end{center}
\caption{\label{Tscaled} Collapse of the shear rate dependence of the kinetic temperature 
using the scaling law, Eq. (\ref{scalingT}).
The kinetic temperature is defined as $T=\langle m_i\delta {\bf v}_i^2\rangle/3$.
In the both panels, the solid lines are proportional to $\gamma^2$, 
which represent the Bagnold's scaling.
The dotted lines are proportional to $\gamma$, which indicate that 
the kinetic temperature is proportional to $\gamma$ above the jamming density.
(a) The linear force model. (b) The Hertzian force model.}
\end{figure}

The scaling laws that are obtained in the present simulation, 
Eqs. (\ref{scalingS}), (\ref{scalingP}), and (\ref{scalingT}), involve several exponents, 
the physical meaning of which is still not clear.
In the rest of this paper, we discuss two important points regarding 
their physical meaning and their universality focusing the criticality of point J.
First, we discuss the model dependence of the exponents in the scaling law for rheology, 
Eq. (\ref{scalingS}), by comparing it with the scaling law obtained 
by Olsson and Teitel (OT) \cite{olsson}.
\begin{equation}
\label{OT}
\frac{\gamma}{S} = |\Phi|^{\beta}f_{\pm}\left(\frac{S}{|\Phi|^{\Delta}}\right), 
\end{equation}
where OT obtain $\Delta=1.2\pm0.1$ and $\beta=1.7\pm0.2$ 
for a two dimensional system with the linear force model.
In order to see the correspondence between Eqs. (\ref{scalingS}) and (\ref{OT}), 
it is more convenient to rewrite Eq. (\ref{OT}) as 
\begin{equation}
\label{OT2}
S = |\Phi|^{\Delta} g_{\pm} \left(\frac{\gamma}{|\Phi|^{\Delta+\beta}}\right).
\end{equation}
Now it is apparent that $\Delta$ is identical to $y_{\Phi}$ in Eq. (\ref{scalingS}), 
and actually $y_{\Phi}$ is $1.2\pm0.1$ for the linear force model in the present three dimensional system.
Therefore, the exponent $\Delta$ (or $y_{\Phi}$) is independent of dimensionality, 
but depends on the force model.
This is plausible when we recall the behavior of the shear modulus with respect to the density is 
independent of dimensionality but depends on the force model \cite{ohern2}.
We also remark that the other exponent $\beta$ is not the same for the two cases: 
$\beta\simeq1.7$ \cite{olsson} and $\beta\simeq0.7$ here.
This discrepancy results from particle dynamics; OT adopt overdamp (massless) dynamics 
whereas the inertia comes into play in the present simulation.
We indeed obtain $\beta\simeq1.7$ for a three dimensional system with overdamp dynamics, 
which is to be presented elsewhere.

The second point we wish to discuss is the relation between the scaling laws 
and the exponents that describe the growing correlation length.
To see this, we rewrite Eq. (\ref{scalingS}) in the following form.
\begin{equation}
\label{hypothesisS}
S(\Phi, \gamma) = \lambda^{-1}S(\lambda^{1/y_{\Phi}}\Phi, \lambda^{1/y_{\gamma}}\gamma),
\end{equation}
where $\lambda$ is an arbitrary nonnegative number.
Here we assume that the correlation length $\xi$ diverges as the system approaches point J; 
i.e., $\Phi\rightarrow0$ and $\gamma\rightarrow0$ (or $T\rightarrow0$).
Then, following the idea of conventional critical phenomena, 
Eq. (\ref{hypothesisS}) implies that $\xi\sim S^{-y_S}$, 
$\xi\sim |\Phi|^{-y_{\Phi}y_S}$, and $\xi\sim\gamma^{-y_{\gamma}y_S}$, 
where $1/y_S$ is the scaling dimension of the shear stress.
These relations indicate that the exponents that involve the algebraic divergence of 
the correlation length are not independent.
Here we only discuss the exponents for the linear force model, 
because few results are known as to the growing correlation length in the Hertzian force model.
Using $\xi\propto|\gamma|^{-0.25}$ that has been confirmed in a sheared granular material 
\cite{hatano2}, we obtain $y_S\sim0.4$ so that $\xi\sim |\Phi|^{-0.5}$.
Note that this exponent $0.5$ agrees with the result of Wyart et al. \cite{wyart}, 
which involves the mean free path of force chains. 
We remark that the same result can be obtained using Eqs. (\ref{scalingP}) and (\ref{scalingT}) 
instead of Eq. (\ref{scalingS}), in which case the scaling dimensions 
of the pressure and the kinetic temperature are approximately $1/0.4$ and $1/0.2$. respectively.


To summarize, we find the scaling laws that describe the shear stress, the pressure, 
and the kinetic temperature near the jamming transition point. 
These scaling laws are related to the exponents that describe the growing correlation length, 
and hence advocate the criticality of point J.
However, the `critical' exponents is found to depend on the force model, 
which indicates the peculiarity of point J in contrast to conventional critical phenomena.

The author gratefully acknowledges helpful discussions with 
Michio Otsuki, Hisao Hayakawa, and Namiko Mitarai.


\begin{thebibliography}{99}
\bibitem{finitesize}
We must be careful that the system size is always larger than the correlation length 
otherwise the finite-size effect comes into play.
In order to advocate the system-size independence of the present result,
we test a larger system, the density of which is close to the critical density;  $\phi=0.645$ and $N=8000$.
We obtain the same data in the range of shear rate that is shown Fig. \ref{shearstress}.
\bibitem{note_phiJ}
Although there is a slight uncertainty in the value of $\phi_J$ ($0.645<\phi_J<0.648$), 
the exponents are insensitive to a specific value used for scaling within this range.
\end{thebibliography}
\end{document}